\documentclass{scrartcl}

\usepackage[sumlimits,intlimits,namelimits]{amsmath}
\usepackage{amssymb}
\usepackage[english]{babel}
\usepackage{bbm}
\usepackage{parskip}
\usepackage{epsfig}

\begin{document}
\bibliographystyle{plain}
\setlength{\parindent}{0cm}
\newif\iffigs\figsfalse
\figstrue

%


\title {Absorption on horizon-wrapped branes}
\author{S.~Ansari$^{a}$\footnote{\tt{saeid.ansari@physik.uni-muenchen.de}}, G.~Panotopoulos$^{a}$\footnote{\tt{grigoris.panotopoulos@physik.uni-muenchen.de}}, I.~Sachs$^{a}$\footnote{\tt{ivo.sachs@physik.uni-muenchen.de}} \\[10pt]
$^{a}$Arnold-Sommerfeld-Center for Theoretical Physics\\
Theresienstra\ss e 37, D-80333 M\"unchen, Germany\\
\\
}
\date{}

\maketitle

\vspace{-250pt}
\hfill{LMU-ASC 1/09}
\vspace{+250pt}

\paragraph{}
\abstract{We compute the absorption cross section of space-time scalars  on a static D2 brane, in global coordinates, wrapped on the $S^2$ of an $AdS_2\times S^2\times CY_3$  geometry. We discuss its relevance for the construction of the dual quantum mechanics of Calabi-Yau black holes.
\clearpage

\section{Introduction}
One of the open problems in gaining an detailed understanding of the AdS\-CFT correspondence is to construct the conformal quantum mechanics dual to $AdS_2$ supergravity backgrounds which arise, for instance, as near horizon geometries of Calabi-Yau black holes with generic D6-D4-D2-D0 charges.

Conformal invariance in quantum mechanics requires the target space of the particle to be itself an $AdS_2$. A natural candidate for a dual theory is thus the quantum mechanics of a set of probe D0- branes in $AdS_2$. Early attempts \cite{Townsend} to describe this quantum mechanics on the Poincar\'e patch of $AdS_2$ have been faced with the complication of a continious spectrum of the Hamiltonian so that there is no well defined ground state degeneracy.

This complication is absent for the Hamiltonian in global coordinates for $AdS_2$ and it has been argued that in the case of a Calabi-Yau black hole with D4- and D0 charge $p^A$ and $q_0$ respectively, the dual quantum mechanics may be that of $q_0$ probe D0-branes on a $AdS_2\times S^2\times CY_3$ target space with $6$-form flux $p^A$. In particular, it was shown\footnote{see also \cite{Mathur,Aspinwall,Douglas} for related discussions.} \cite{StromingerBHQM} that the asymptotic degeneracy of collective excitations of the probe D0-branes described by a gas of D2 branes on $S^2$ with total D0 charge $q_0$ agrees with the Beckenstein-Hawking entropy formula. 
As usually in string , black holes the large degeneracy required to reproduce the black hole entropy comes from the internal dimensions, in this case the Landau levels of the D2 branes in the $6$-form flux on the CY. Of course, for this type of black holes the dual 1+1 dimensional CFT is known in eleven dimensions \cite{MSW}. Nevertheless the construction of the dual quantum mechanics in ten dimensions is an important problem in particular in view of the CY-black holes with D6-charge.

Apart from reproducing the entropy of the black hole one expects that the dual quantum mechanics should provide a microscopic description of the absorption- and Hawking emission of space-time fields by the black hole. 
In this paper we address the problem of absorption of space-time fields by the world volume of collective D2 branes with generic D0 probe-brane charge.
For large probe D0-brane charge the coupling of the space-time fields to the world-volume quantum mechanics becomes large
so that linearized perturbation theory is no longer applicable and back-recation on geometry has to be taken into account.
In the calculations presented here we will thus restrict ourselves to D2 branes with small D0-charge so that perturbation theory is applicable. With this restriction the 2-branes are confined to the near horizon regime. In terms of the proposed $AdS_2/QM$ correspondence a single horizon wrapped D2 brane with small probe D0 charge could possibly be interpreted as the dual description of a small, non-extremal perturbation of the black hole background.

We find that the quantum mechanical absorption cross section seen by an observer static in asymptotic (Poincar\'e) time vanishes linearly in $\omega$ for small frequencies for small but non-zero probe D0 charge and has non-analytic ($\omega\log(\omega)$) behavior for vanishing  probe D0 charge. This is in disagreement with the classical s-wave absorption cross-section by the black hole which vanishes quadratically in $\omega$ for small frequencies \footnote{See also Appendix 1}\cite{Gibbons}.

In section 2 we start with a brief review of the proposed conformal quantum mechanics for D4-D0 Calabi-Yau black holes. In section 3 we compute the cross section for the absorption of dilatons on the two-brane.  We conclude in section 4. Some technical details of the calculation are contained in the appendices.


\section{The D2-brane conformal quantum mechanics }
We consider a quantum mechanical system of non-interacting probe D0-branes with $AdS_2\times S^2\times CY_3$ target space. This background can be obtained, for instance, as a compactification of IIA theory on a CY-manifold X, with $q_0$ D0-branes and D4-branes wrapped on a four-cycle ${\cal{P}}$ in the homology class $p^A\Sigma_A$. The four-cycle has to be holomorphic for the configuration to be BPS. The radius of $AdS_2$ and $S^2$ depends on the charges as $R=\sqrt{2}(Dq_0)^{1/4}$ \cite{StromingerSB}.
For the SUGRA description to be valid we need to assume that
\begin{equation}\label{hir}
|q_0|>\!\!>p^A>\!\!>0\ .
\end{equation}
\subsection{BPS Branes}
It has been shown in \cite{StromingerSB} that this geometry admits supersymmetric probe D2 branes, wrapped on $S^2$ with arbitrary D0-charge bound to them. The branes are static in global coordinates of the $AdS_2$. Their radial position is determined in terms of the D0-charge 
bound to them. 
Furthermore, there is a large Landau level degeneracy, proportional to $D=D_{ABC}p^Ap^Bp^C$ due to the magnetic coupling to the D4 flux in the CY.
It has been argued in \cite{StromingerBHQM} that these D2 branes should be interpreted as collective excitations in the quantum mechanics of probe D0-branes $AdS_2\times S^2\times CY_3$. In particular, the asymptotic growth of such states with large D0-charge correctly reproduces the Beckenstein-Hawking entropy of the 4 dimensional black hole with equal charges \cite{StromingerBHQM}.

Here we will be interested in the absorption cross section on such D2-branes as seen by an  asymptotic observer. For this we need the coupling of the 2-branes to the space-time fields. This can be inferred from the Born-Infeld action
\begin{equation}\label{10}
S=-\mu_2\int d^3\xi e^{-\Phi}\sqrt{-|G+2\pi\alpha' F|}+\mu_2\int_{\Sigma_3}[P[C^{[3]}]+2\pi\alpha'F\wedge P[ C^{[1]}]]\,,
\end{equation}
where $G_{ab}$ is the induced string frame world volume metric for a given 10-dimensional string metric and $C^{[1]}$ is the RR 1-form in IIA theory with
\begin{equation}
dC^{[1]} = \frac{R^2}{q_0} \cosh \chi d\tau \wedge d\chi\ .
\end{equation}
Furthermore, $F_{ab}$ is the field strength of the world-volume $U(1)$ gauge potential, $A$ with background value
\begin{equation}
A=-\frac{f}{2\pi\alpha'}\cos(\theta)d\phi\ .
\end{equation}
Finally the background value of the dilaton is given by
\begin{equation}
e^{\Phi_0}=\frac{q_0}{R}\,.
\end{equation}
\subsection{Vibration Modes}
In what follows we will work in the static gauge ($\xi^a \equiv X^a$) and we will neglect internal excitations  levels in X which are suppressed in the approximation (\ref{hir}). In this case there is exactly one transverse scalar field parametrizing the radial position of the brane in  $AdS_2$ as well a gauge field $A_a$ which is again equivalent to a scalar field $\psi$ on-shell. The fluctuations $\delta\chi$ and $\psi$ are functions of $(\tau,\theta,\phi)$.

We take the $D0$-charge on the 2-brane, $\tilde q_0\propto\int_{S^2}F$ to be fixed.
In this case the quantum mechanical (s-wave) excitation $\delta\chi(t)$ decouples from all other excitations and can thus be treated separately. Upon substitution of the $AdS_2\times S^2$ metric
\begin{equation}
ds^2=R^2[-\cosh^2\chi d\tau^2 + \chi^2]
\end{equation}
in the Born-Imfeld action and expanding up to second order in derivatives we find the action for the transverse scalar $\chi$
\begin{equation}
S_{D2}= \frac{1}{g^2}\int \left[\frac{1}{2}\dot\chi^2 -V_1(\chi)+V_2(\chi)\right]d\tau\,,
\end{equation}
with
\begin{equation}
V_1(\chi)=\frac{\sqrt{M_2^2+M_0^2}}{M_2}\cosh(\chi)\ ,\qquad V_2(\chi)=\frac{M_0}{M_2}\sinh(\chi)\,,
\end{equation}
and $g^{-2}=M_2 Re^{-\Phi_0}=16\pi\mu_2D$, $M_0= 4\pi\mu_2f$, $M_2=4\pi\mu_2R^2$.  $V_2(\chi)$ originates from the Chern-Simons term in (\ref{10}). The probe brane potential has a minimum at $\sinh(\chi_0)=\frac{M_0}{M_2}$. In Poincar\'e coordinates this brane oscillates in and out of the the horizon as can be seen in the Carter-Penrose diagram in figure 1.
\begin{figure}
\begin{center}
\epsfig{file=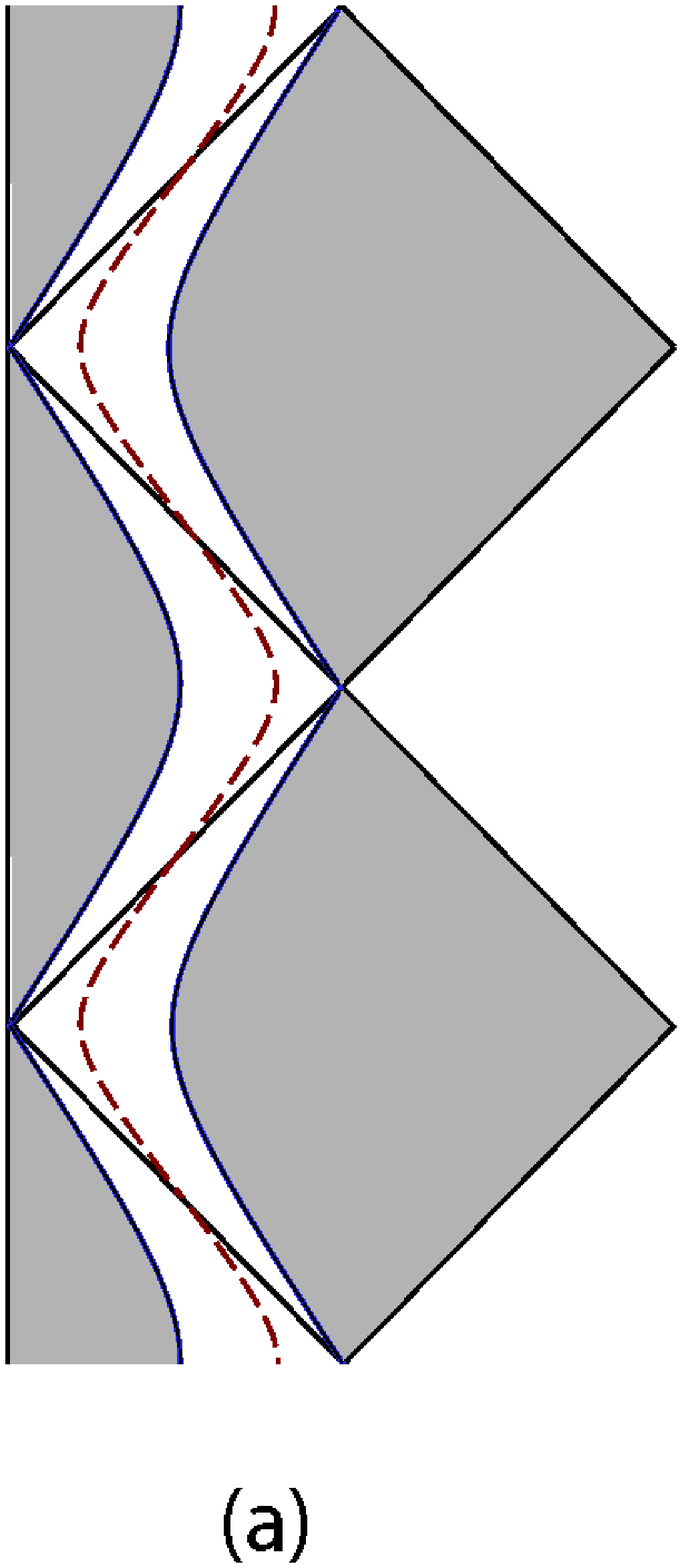,height=8cm} \hfil
\epsfig{file=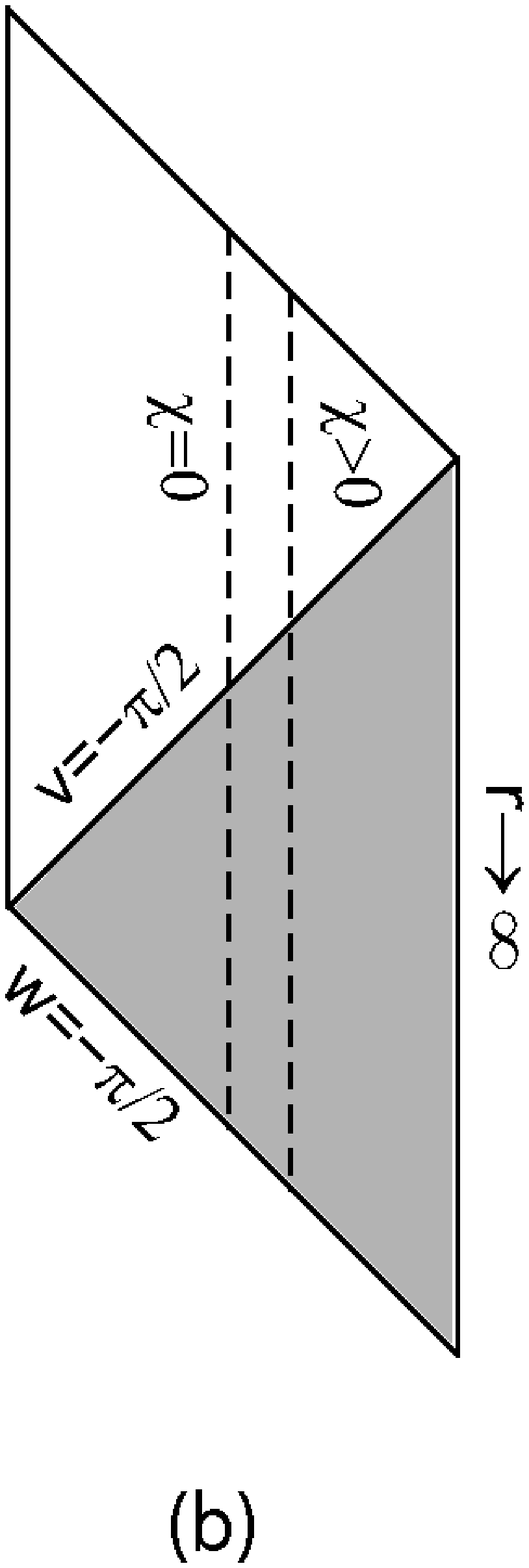,height=8cm}
\end{center}
\caption{(a) Penrose diagram for brane trajectory at near horizon in Poincar\'e coordinates view. The dashed line shows the brane position which oscillates around the horizon. (b) Brane in global AdS, dashed lines present the brane position. }
\end{figure}

Upon expanding the potential $V$ to second order in $\delta\chi$ we obtain a harmonic oscillator with frequency $1$ (in units of $1/R$) which is quantized in the usual way as 
\begin{equation}\label{ho}
\delta\chi(\tau)=\frac{g}{\sqrt{2}}\left(e^{i\tau} a+e^{-i\tau}a^\dagger\right)\,,
\end{equation}
where $a$ and $a^\dagger$ are the annihilation- and creation operators for the harmonic oscillator.

Next we consider quadratic fluctuations with non-vanishing angular momentum. Upon expanding the action (\ref{10}) to second order in the fluctuation in the position $\delta \chi$ and the gauge field, $f_{ab}=\partial_a A_b -\partial_b A_a$ we now get
\begin{eqnarray}\label{S2}
S^{(2)}&=&\frac{1}{ g^2 4\pi}\int d^3\xi\sqrt{-\tilde g}\left[-\frac{1}{2}(\delta\chi)^2-
\frac{1}{2}
\tilde g^{ab}\partial_a\delta \chi\partial_b\delta \chi\right.\\
&&\left.\qquad\qquad-
\frac{1}{4}\frac{(2\pi\alpha')^2}{R^4+f^2} \tilde g^{ac}\tilde g^{bd}f_{ab}f_{cd}+\frac{1}{2}\frac{ (2\pi\alpha')^2}{f\sqrt{R^4+f^2}}F^{ab}f_{ab}\delta \chi\right]\nonumber\,,
\end{eqnarray}
where
\begin{equation}
\tilde g_{ab}=\begin{pmatrix}-1&0&0\cr 0&1&0\cr0&0&\sin^2(\theta)  \end{pmatrix}\,.
\end{equation}
Thus the dynamics of the quadratic fluctuations on a 2-brane in $AdS_2 \times S^2$ wrapped on $S^2$ is identical to that of a brane fluctuating in ${\bf R}^{1,1} \times S^2$  with a non-trivial potential
$V(\delta\chi,f_{ab})$. This coupling persists in the absence of a $D0$ probe brane charge as a consequence of the Chern-Simons term in (\ref{10}). The equations of motion for the fluctuations obtained from (\ref{S2}) are then found to be
\begin{eqnarray}\label{ema}
\frac{\delta S^{(2)}}{\delta A_e}&=&(2\pi\alpha')^2\tilde \nabla_a \left(f^{ae}-\frac{\sqrt{R^4+f^2}}{f}F^{ae}\delta\chi\right)=0\,,\\
\frac{\delta S^{(2)}}{\delta(\delta \chi)}&=&\left(\tilde \nabla^2-1\right)\delta\chi +\frac{1}{2}\frac{(2\pi \alpha')^2}{f\sqrt{R^4+f^2}}F^{ab}f_{ab}=0\,,\nonumber
\end{eqnarray}
where the indices are now lifted with $\tilde g$ and $ \tilde \nabla^2$ is with respect to $\tilde g$.
To continue, using  (\ref{ema}), we express $f_{ab}$ in terms of a scalar field through
\begin{eqnarray}\label{psi}
f^{ae}=\frac{R^2}{(2\pi\alpha')}\frac{1}{\sqrt{\tilde g}}\epsilon^{aeg} \partial_g\psi+\frac{\sqrt{R^4+f^2}}{f}F^{ae}\delta\chi\,.
\end{eqnarray}
Note that this change of variables is valid on-shell only. In terms of these new fields the equations of motions then take the form
\begin{eqnarray}
\tilde \nabla^2\psi-\frac{\sqrt{R^4+f^2}}{R^2}\partial_\tau \delta\chi &=&0\,,\\
\tilde \nabla^2 \delta\chi+\frac{R^2}{\sqrt{R^4+f^2}}\partial_\tau\psi&=&0\,.\nonumber
\end{eqnarray}
In order to diagonalize this system we expand $\psi$ and $\delta \chi$ in spherical harmonics as $f(\tau,\theta,\phi)=f_{lm}e^{-i\Omega\tau}Y_{lm}(\theta,\phi)$. This leads to
\begin{equation}\label{mat}
\begin{pmatrix} \frac{\sqrt{R^4+f^2}}{R^2} \left(-l(l+1)+\Omega^2\right)&-i\Omega\cr i\Omega&\frac{R^2}{\sqrt{R^4+f^2}}\left(-l(l+1)+\Omega^2\right)\end{pmatrix}
\begin{pmatrix} \delta\chi_{lm}\cr \psi_{lm}\end{pmatrix} =0\,.
\end{equation}
The eigenfrequencies are given by $\Omega= l$ and $\Omega=l+1$ with corresponding eigenmodes
\begin{eqnarray}\label{vec}
\Phi^{0,1}=\frac{1}{\sqrt{2}}\begin{pmatrix}1\cr\pm i\frac{\sqrt{R^4+f^2}}{R^2}\end{pmatrix}\varphi^{0,1}_{lm}\,,
\end{eqnarray}
respectively, where $\varphi^{0,1}_{lm}$ are canonically normalized free fields with dispersion relation $\Omega= l$ for $\varphi^{0}_{lm}$ and $\Omega=l+1$ for  $\varphi^{1}_{lm}$. We can thus quantize $\delta\chi$ and $\psi$ in terms of free fields $a_{lm}$ and $b_{lm}$ as\footnote{We use the convention where $\int_{S^2}Y_{lm}Y_{l'm'}^*=4\pi\delta_{ll'}\delta_{mm'}$.} 
\begin{eqnarray}\label{chipsi}
\begin{pmatrix}\delta\chi_{lm}(\tau)\cr\psi_{lm}(\tau)\end{pmatrix}&=&\frac{g}{2}\left[
\frac{a_{lm}}{\sqrt{l}}e^{il\tau}Y_{lm}\begin{pmatrix}1\cr-i\frac{\sqrt{R^4+f^2}}{R^2}\end{pmatrix}+\frac{a^\dagger_{lm}}{\sqrt{l}}e^{-il\tau}Y^*_{lm}\begin{pmatrix}1\cr i\frac{\sqrt{R^4+f^2}}{R^2}\end{pmatrix}\right.\nonumber\\
&&\qquad\left.+\frac{b_{lm}}{\sqrt{l+1}}e^{i(l+1)\tau}Y_{lm}\begin{pmatrix}1\cr i\frac{\sqrt{R^4+f^2}}{R^2}\end{pmatrix}\right.\\\nonumber&&\qquad\left.+\frac{b^\dagger_{lm}}{\sqrt{l+1}}e^{-i(l+1)\tau}Y^*_{lm}\begin{pmatrix}1\cr- i\frac{\sqrt{R^4+f^2}}{R^2}\end{pmatrix}\right]\,.\nonumber
\end{eqnarray}
The integer valuedness of the spectrum is to be expected since in global coordinates time $\tau\simeq \tau+2\pi$ is compactified. In Poincar\'e time this means that $t=-\infty$ and $t=\infty $ are identified.

\section{Absorption }
The dilaton and the volume moduli of the CY (the latter is a fixed scalar) couple to the radial position $\chi$ of the 2-branes in $AdS_2$ as well as to the world-volume gauge potential  through the DBI-term in (\ref{10}). The RR 1-form field couples to the world-volume gauge potential and the radial position through the CS-term in (\ref{10}). We will focus on the dilaton absorption at present.  To begin with we consider the quantum mechanical (s-wave) absorption of a dilaton $\delta\phi$, which then couples to the transverse position as
\begin{equation}\label{a1}
S_{D2}= \frac{1}{g^2}\int \;(-\delta\phi)\left(\frac{1}{2}\dot\chi^2 -V_1(\chi)\right)d\tau\,.
\end{equation}
The potential $V_2$ is induced by the CS term in (\ref{10}) and thus does not couple to the dilaton. 
To continue we need to distinguish between small- and large D0 probe-brane charge 
since $\delta\phi$ does not couple to $V_2$. For $M_0<\!\!<M_2$ we have $V_1(\chi_0)\simeq 1$ so that the back-reaction of the probe brane on the dilaton $\Phi$ can be neglected. For $M_2<\!\!<M_0$ on the other hand $V_1(\chi_0)\propto \frac{f^2}{R^4}>\!\!>1$ so that in the linearized approximation back reaction of the probe brane destabilizes the supergravity background\footnote{Of course, in the full non-linear theory a (constant) deformation of the dilaton is a marginal deformation since the position of the probe brane is a smooth function of the string coupling.}. We will thus not consider this possibility. On the other hand, for small $D0$ probe brane charge $\tilde q_0$, the probe brane trajectory is within the near horizon region of the Poincar\'e patch of $AdS_2$.
In this case it is interesting to compute the absorption cross section as observed by an asymptotic observer and compare it to the classical black hole absorption cross section.
\subsection{Spherical Excitations without D0-charge}
We will be interested in finding the absorption cross section seen by an observer static in asymptotic (Poincar\'e) time, $t$. We first consider the case of a probe brane without $D0$-charge, $M_0=0$. 
In the near horizon $AdS_2$ the classical solutions for an s-wave dilaton perturbation with frequency $\omega$  with respect to the Poincar\'e time are given by
\begin{equation}
\delta\phi_\omega(t)= e^{ i \omega(t\mp\frac{R^2}{r})}\ .
\end{equation}

\subsubsection{Quantum mechanical mode}
Aa a warm-up we first treat obtain the absorption cross section in the harmonic oscillator approximation (\ref{a1}), ie. assuming vanishing angular momentum of the excitations on the probe-brane.  In this case the coupling of the dilaton perturbation to the transverse excitation of the 2-brane is given by
\begin{eqnarray}
S_{int}&=&\frac{1}{2g^2}\int{\delta\phi}\left(- \dot{\delta\chi}^2+\delta\chi^2\right)\\
&=&\frac{1}{g^2}\int \;\delta\phi  \delta\chi^2 d\tau  \nonumber\,,
\end{eqnarray}
where "$\;\dot{}\;$" indicates derivative w.r.t. global time $\tau$. 
To leading order in $g$ we get for a ingoing dilaton upon inserting (\ref{ho})
\begin{equation}
\langle2|S_{int}|0\rangle=\frac{\sqrt{2}}{2}\int\limits_{-\frac{\pi}{2}}^{\frac{3\pi}{2}} e^{ i \omega(t-\frac{R^2}{r})} e^{-2i\tau}d\tau\,.
\end{equation}
Note that for $M_0=0$ there is no transition from the ground state to  the first excited state $|1\rangle$. 
In order to evaluate the integral we change to Kruskal coordinates which for $\chi_0=0$ are given by
\begin{eqnarray}
vR&=&t-\frac{R^2}{r}=R\tan(\frac{\tau-\frac{\pi}{2}}{2})\,,\\
wR&=&t+\frac{R^2}{r}=R\tan(\frac{-\tau+\frac{\pi}{2}}{2})\ .\nonumber
\end{eqnarray}
Then
\begin{eqnarray}
\langle2|S_{int}|0\rangle&=&-\sqrt{2}\int\limits_{-\infty+i\epsilon}^{\infty+i\epsilon}  \frac{1}{1+v^2}e^{ i \omega R v} \left(\frac{1-iv}{1+iv}\right)^2d v  \label{c1}\\
&=&2\sqrt{2}\pi R\omega (R\omega-1) e^{-R\omega}\,,\nonumber
\end{eqnarray}
where the contour $C$ closes in the upper half plane for $\omega>0$ and passes above the pole at the origin in accordance with the $i\epsilon$ prescription for absorption.  The boundary term in (\ref{c1}) ensures the correct fall-off at $v=\pm \infty$ required in order to recast this amplitude as a contour integral.
Excitation to higher level is also possible and the corresponding amplitude can be calculated by expanding $V_1$ to higher order in $\chi$. The corresponding amplitude is sub-leading in $g$.

To determine the cross section for an s-wave dilaton into an s-wave excitation on the brane for an asymptotic observer in  $AdS_2\times S^2$ geometry we note that the ingoing flux of the field $\phi$ is given by\footnote{We have ${\cal{F}}=4\pi\frac{1}{2i}\sqrt{g} g^{rr} \left(\bar \phi\partial_r\phi-   \phi\partial_r\bar \phi\right)$ with $ds^2=-\frac{r^2}{R^2}dt^2+ \frac{R^2}{r^2} dr^2 +R^2 d\Omega_2^2$. The $4\pi$ comes from integration over the $S^2$.}  ${\cal{F}}_{AdS_2}=4\pi\omega$.  The $AdS_2$ cross section for this process is thus given by ($T=2\pi R$)
\begin{eqnarray}
\sigma_{AdS_2}&=&\frac{|{\cal{A}}_{0\to 2}|^2}{T {\cal{F}}}\,.\\
&=&R\omega(R\omega-1)^2e^{-2\omega R}\nonumber
\end{eqnarray}
This is only part of the complete absorption cross section for an s-wave dilaton, since we ignored higher angular momentum excitation so far. Never the less this partial cross section allows us to discuss some qualitative features. First we note that in spite of the discreteness of the spectrum of the D2-brane Hamiltonian, the D2-brane can absorb arbitrarily small frequencies with respect to Poincar\'e time. This is in agreement with the classical picture. On the other hand, we can compare this cross section with the classical s-wave absorption cross section, proportional to the square of transmition coefficient which has a universal form for small frequencies \cite{Gibbons} given by $|T|^2 \simeq (R\omega)^2$. For a small non-extremal perturbation of the black hole mass that could be equivalent to adding a wrapped neutral D2-brane, we thus have $\delta|T|^2 \simeq R\delta R\omega^2$ which vanishes like $\omega^2$.  We thus conclude that classical s-wave absorption cross section for an asymptotically flat black hole background is not reproduced by the microscopic cross section of the wrapped D2-brane.

\subsubsection{Higher partial waves}
Recalling that the Chern-Simons term does not couple to $\delta\phi$ we get the following interaction  including all angular momenta on $S^2$:
\begin{eqnarray}\label{hpw}
S_{int}=\frac{1}{2g^24\pi}\int\limits_{-\frac{\pi}{2}}^{\frac{3\pi}{2}}\int_{S^2}\;\delta\phi(t)\bigg[ && \!\!\!\!\!\!\!\!\!\!\!\sqrt{-\tilde g}\left( \tilde g^{ab}\partial_a\delta\chi\partial_b\delta\chi- \frac{R^4}{R^4+f^2} \tilde g^{ab}\partial_a\delta\psi\partial_b\delta\psi \right) \nonumber\\
&&+2\frac{R^2}{\sqrt{R^4+f^2}}\delta\chi\partial_\tau \psi\bigg]d\tau\,.
\end{eqnarray}
Using (\ref{psi}) and (\ref{chipsi}) we then end up with
\begin{eqnarray}\label{cao}
S_{int}=\frac{1}{4}\sum\limits_{l\neq 0,m}\int\limits_{-\frac{\pi}{2}}^{\frac{3\pi}{2}}&&\!\!\!\!\!\!\!\!\!\!\!e^{ i \omega(t-\frac{R^2}{r})} (-1)^m\left[\frac{1}{l}a^\dagger_{lm}a^\dagger_{l-m}e^{-i2l\tau}\right.\\&&\left. \!\!\!\!\!\!\!\!\!\!\!\!\!\!\!\!\!\!\!\!\!\! +
\frac{1}{\sqrt{l(l+1)}}a^\dagger_{lm}b^\dagger_{l-m}e^{-i(2l+1)\tau}\
+
\frac{1}{(l+1)}b^\dagger_{lm}b^\dagger_{l-m}e^{-i2(l+1)\tau}\right]d\tau\,.\nonumber
\end{eqnarray}
Here we have ignored terms that contain the annihilation operators $a_{lm}$ and $b_{lm}$ since we take the initial state to be the ground state so that they do not contribute to the absorption amplitude. Note that if we include the $l=0$ mode for the last term then we recover the quantum mechanical mode discussed before.  The transition amplitudes for fixed $l>0$ and $m$  in leading order are then found to be
\begin{eqnarray}\label{aa}
\langle a,l~m;a,l~-m|S_{int}|0\rangle&=&\frac{c_m (-1)^m}{l}\int\limits_{-\frac{\pi}{2}}^{\frac{3\pi}{2}}e^{ i \omega(t-\frac{R^2}{r})}e^{-i2l\tau} d\tau\\
&=&4\pi c_m\frac{(-1)^{l+m+1}}{l} M_{2l,\frac{1}{2}}(2R\omega) \,,\nonumber
\end{eqnarray}
with $c_m=1/2$ for $m\neq 0$, $c_m=\sqrt{2}/4$ for $m=0$ and $\langle a,l~m;a,l~-m|$ denotes the final state consisting of two excitations of type $a$ with angular momentum $l$ and $L_3=\pm  m$ respectively. Furthermore
\begin{equation}
M_{\lambda,\mu}(z)=\frac{z^{\mu+\frac{1}{2}}}{2^{2\mu}B(\frac{1}{2}+\mu+\lambda,\frac{1}{2}+\mu-\lambda)}\int\limits_{-1}^1e^{\frac{1}{2}zt}(1+t)^{\mu-\lambda-\frac{1}{2}}(1-t)^{\mu+\lambda-\frac{1}{2}} d t \;
\end{equation}
is the Whittacker function \cite{Gradstein}. Similarly
\begin{eqnarray}\label{bb}
\langle b,lm;b,l~-m|S_{int}|0\rangle
&=&4\pi c_m\frac{(-1)^{l+m}}{l+1} M_{2l,\frac{1}{2}}(2R\omega) \,,
\end{eqnarray}
for two b-type excitations in the final state and
\begin{eqnarray}\label{ab}
\langle b,lm;a,l~-m|S_{int}|0\rangle
&=&\pi i\frac{(-1)^{l+m+1}}{\sqrt{l(l+1)}} M_{2l-1,\frac{1}{2}}(2R\omega) \,,
\end{eqnarray}
for one a-type and one b-type excitation in the final state. In order to obtain the total  cross section for absorption of an s-wave dilaton on a 2-brane without D0-charge we have to sum the partial cross section over $l$ and $m$. Note that taking into account the quantum mechanical mode the sum over $l$ starts from $1$ for (\ref{aa}) and (\ref{ab}) and from 0 for (\ref{bb}). Furthermore $m=0,\cdots,l$ for (\ref{aa})  and (\ref{bb}) due to symmetry under $m\to -m$, and  $m=-l,\cdots,l$(\ref{ab}). Putting all this together and dividing by the incomming flux we end up with
\begin{eqnarray}\label{fs}
\sigma_{AdS_2}&=&\frac{1}{4R\omega} \sum\limits_{l=1}^\infty\frac{1}{l}|M_{2l,\frac{1}{2}}(2R\omega)|^2
+\frac{1}{32R\omega}\sum\limits_{l=1}^\infty\frac{2l+1}{l(l+1)}|M_{2l-1,\frac{1}{2}}(2R\omega) |^2\,.\nonumber\\
\end{eqnarray}
While we are not aware of any closed expression for the above sum we can never the less extract the low frequency behavior with the help of an integral approximation of the sums in (\ref{fs}) (see Appendix 2) leading to
\begin{eqnarray}\label{fsl}
\sigma_{AdS_2}&\simeq&-5\;R\omega(\log(R\omega)+const)\;, \qquad \omega\to 0\,.
\end{eqnarray}
The total absorption cross section is thus non-analytic at $\omega=0$. The absorption cross section for space-time scalars with non-vanishing angular momentum can be obtained along the same lines. We indicate the modifications in Appendix 3.

\subsection{Spherical Excitations with D0-charge}
Let us now consider a charged D2 brane with $0<M_0<< M_2$. In this case the interaction term at leading order in $g$ is linear. in particular, for the absorption of an s-wave dilaton perturbation
\begin{eqnarray}
S_{int}
&=& \frac{M_0}{M_2g^2}\int\limits_{-\frac{\pi}{2}}^{\frac{3\pi}{2}}\;\delta\phi(t) \cosh(\chi_0)(\chi-\chi_0)d\tau\,,
\end{eqnarray}
so that leading order absorption amplitude becomes
\begin{eqnarray}\label{d00}
\langle1|S_{int}|0\rangle&=&\frac{M_0\cosh(\chi_0)}{\sqrt{2}gM_2}\int\limits_{-\frac{\pi}{2}}^{\frac{3\pi}{2}}  e^{ i \omega(t-\frac{1}{r})} e^{-i\tau}d\tau\,. 
\end{eqnarray}
This integral can again be obtained in closed form with a suitable transformation of variables (see Appendix 4) leading to
\begin{eqnarray}
\langle1|S_{int}|0\rangle&=&-i\frac{M_0\cosh(\chi_0)}{gM_2}2\sqrt{2}\pi  R\omega e^{-R\omega}\\
&\simeq&-i\frac{f}{gR^2}2\sqrt{2}\pi  R\omega e^{-R\omega}\,,\nonumber
\end{eqnarray}
where we have assumed $f/R^2<\!\!<1$. Proceeding as above we then obtain the $AdS$ cross section
\begin{eqnarray}
\sigma_{AdS_2}&=&\frac{f^2}{g^2R^4}R\omega e^{-2\omega R}\,.
\end{eqnarray}

\subsubsection{Dilaton with arbitrary angular momentum}
We now consider the absorption of a dilaton with arbitrary angular momentum on a probe  brane with finite D0-charge, $f$. Concretely we take the dilaton perturbation of the form

\begin{equation}
\delta\phi(t,\theta,\varphi)= Y_{lm}(\theta,\varphi)e^{ i \omega(t\mp\frac{R^2}{r})}\ .
\end{equation}
For $f\neq 0$ the dilaton couples to the probe brane oscillation through the first order interaction
\begin{eqnarray}
S_{int}&=&\frac{f}{R^2g^24\pi}\int\limits_{-\frac{\pi}{2}}^{\frac{3\pi}{2}}\int_{S^2}\sqrt{-\tilde g}\;\delta\phi\left[ \cosh(\chi_0)\delta\chi +\frac{(2\pi\alpha')^2}{2f^2}\sinh(\chi_0)F^{ab}  f_{ab}\right]d\tau
\nonumber\\
&=&\frac{f}{g^2 R^2\pi}\int\limits_{-\frac{\pi}{2}}^{\frac{3\pi}{2}}\int_{S^2}\sqrt{-\tilde g}\;\delta\phi(t,\theta,\varphi)\left[2\frac{\sqrt{R^4+f^2}}{R^2}\delta\chi +\partial_0\psi\right]d\tau\,.
\end{eqnarray}
The leading order, non-vanishing components of the transition amplitude for absorption of a dilaton with angular momentum $l,m$ into an a-type excitation is thus given by
\begin{eqnarray}
\langle a;lm|S^{lm}_{int}|0\rangle&=&\sqrt{R^4+f^2}\frac{f}{2g R^4\sqrt{l}}\int\limits_{-\frac{\pi}{2}}^{\frac{3\pi}{2}}  e^{ i \omega(t-\frac{R^2}{r})}(2+l) e^{-il\tau}d\tau \nonumber\\
&\simeq&- \frac{f(2+l)i^l}{gR^2\sqrt{l}}\pi M_{l,\frac{1}{2}}(2R\omega)\,,
\end{eqnarray}
where we have ignored terms of order $f^2$ and higher when going from the first to the second line. Indeed we have argued above that for charged probe branes the absorption cannot be treated perturbatively unless $f<\!\!<R^2$.   Similarly we get for absorption into an b-type excitation
\begin{eqnarray}\label{db}
\langle b;lm|S^{lm}_{int}|0\rangle
&=&-  \frac{f(1-l)i^{l+1}}{gR^2\sqrt{l+1}}\pi M_{l-1,\frac{1}{2}}(2R\omega) +O(f^2)\,.
\end{eqnarray}
Note that a naive application of (\ref{db}) for $l=0$ leads to a $\sqrt{2}$ discrepancy with the earlier result (\ref{d00}). This apparent contradiction is resolved by recalling that for $l=0$ there is no $a$-mode so that there is an extra $\sqrt{2}$ in the normalization of (\ref{chipsi}). Note also that for $l=1$ the amplitude for absorption in to a $b$-type excitation vanishes.

The total cross section for the absorption of $\phi_{am}$ is the again obtained by adding the squares of the a- and b-type amplitudes (without summing over m). This gives
\begin{eqnarray}\label{ggg}
\sigma_{AdS_2}&=&\frac{f^2}{8gR\omega R^4} \left[\frac{(2+l)^2}{ l} |M_{2l,\frac{1}{2}}(2R\omega)|^2
+\frac{(1-l)^2}{(l+1)}|M_{l-1,\frac{1}{2}}(2R\omega) |^2\right]\,,\nonumber\\
\end{eqnarray}
which vanishes linearly for small $\omega$.

\section{Conclusions}
We have obtained analytic expressions for the absorption cross section of space-time scalars on horizon wrapped D2-branes, static in global coordinates of the near horizon $AdS_2$ geometry. The fact that these amplitudes can be computed exactly may come as a surprise since the probe 2-brane describes a complicated trajectory in the asymptotic Poincar\'e coordinates.

An interesting feature is that although the Hamiltonian of the D2-brane has a discrete spectrum with spacing given by the inverse of the radius of the horizon the D2 brane can absorb arbitrarily small frequencies with respect o an asymptotic observer.

In view of a possible interpretation for a dual interpretation of 4 dimensional CY-black holes in terms of the quantum mechanics of probe D2-branes wrapped on the $S^2$ of their near horizon geometry an encouraging result would have been to find agreement for the low energy absorption cross section on both sides. Our concrete calculation shows however that this is not case since the microscopic asorption  cross section on the 2-brane does not have the correct behavior at small frequencies compared to the classical absorption cross section of massless scalars which vanishes  quadratically in $\omega$.

We should mention that we only considered the bosonic sector of the world volume theory. However, it is not hard to see that fermions give a vanishing contribution at the lowest (quadratic) level. 
Also we have not considered fixed scalars in this paper although their inclusion should be straight forward.

\paragraph{Acknowledgments:}
The authors would like to thank
J. D. Laenge 
for helpful discussion. This work was supported in parts by SPP-1096, the Transregional Collaborative Research Centre TRR 33 and the  Excellence cluster "Origin and Structure of the Universe" of the DFG  as well as the Virtual Institute for Particle Cosmology funded by the Helmholtz Society. 
\section{Appendix 1}
For completeness we include the computation of the absorption cross section for a massless scalar field in the background of a four-dimensional extremal Reissner-Nordstrom black hole.
For this we use well-known techniques, namely we need to solve the wave equation (radial part) of the scalar field in the far region regime as well as in the near-horizon regime,
then match the solutions
and finally compute the absorption coefficient. If $R$ is the reflection coefficient and $T$ the transition coefficient, then the absorption cross section is given by the optical theorem for
the four-dimensional case
\begin{eqnarray}
\sigma_l & = & \frac{\pi (2l+1)}{\omega^2} |T|^2 \,,\\
1 & = & |T|^2+|R|^2\,,\nonumber
\end{eqnarray}
where $l$ is the angular momentum and $\omega$ is the frequency. In the far region regime ($r \rightarrow \infty$) the solution of the radial part is given in terms of the usual (cylindrical) Bessel functions
\begin{equation}
R_{far}(r)=\frac{1}{\sqrt{r}} (\alpha J_{l+\frac{1}{2}}(\omega r)+\beta J_{-l-\frac{1}{2}}(\omega r))\,.
\end{equation}
In the near horizon regime ($r \simeq r_H$) it is convenient to write the black hole solution in the form
\begin{equation}
ds^2=-\frac{1}{y^2}dt^2+\frac{r_H^2}{y^2}dr^2+r_H^2 d \Omega_2^2\,,
\end{equation}
where $y=r_H/z$ and $z=r-r_H$. Then the solution of the ``radial part'' is given by
\begin{equation}
R_{Hor}(y)=\sqrt{y} (C_1 J_{l+\frac{1}{2}}(\omega r_H y)+J_{-l-\frac{1}{2}}(\omega r_H y))\,,
\end{equation}
or
\begin{equation}
R_{Hor}(r)=\frac{1}{\sqrt{r-r_H}} (A J_{l+\frac{1}{2}}(\frac{\omega r_H^2}{r-r_H})+B J_{-l-\frac{1}{2}}(\frac{\omega r_H^2}{r-r_H}))\,.
\end{equation}
Matching the solutions we obtain the coefficients $\alpha, \beta$ is terms of $A,B$ as follows
\begin{eqnarray}
\frac{\alpha}{\Gamma(l+\frac{3}{2})} \left ( \frac{\omega}{2} \right )^{l+\frac{1}{2}} & = & \frac{B}{\Gamma(-l+\frac{1}{2})} \left ( \frac{\omega r_H^2}{2} \right )^{-l-\frac{1}{2}} \,,\\
\frac{A}{\Gamma(l+\frac{3}{2})} \left ( \frac{\omega r_H^2}{2} \right )^{l+\frac{1}{2}} & = & \frac{\beta}{\Gamma(-l+\frac{1}{2})} \left ( \frac{\omega}{2} \right )^{-l-\frac{1}{2}}\,,\nonumber
\end{eqnarray}
where $\Gamma(x)$ is the usual Gamma function. Finally the reflection coefficient is given by
\begin{equation}
R=\frac{\left ( \frac{\Gamma(l+\frac{3}{2})}{\Gamma(-l+\frac{1}{2})}  \right )^2 \left ( \frac{4}{\omega^2 r_H^2} \right )^{2l+1}-1}{\left ( \frac{\Gamma(l+\frac{3}{2})}{\Gamma(-l+\frac{1}{2})}  \right )^2 \left ( \frac{4}{\omega^2 r_H^2} \right )^{2l+1}+1}\,.
\end{equation}
\section{Appendix 2}
In order to isolate the low frequency behavior of the cross section (\ref{fs}) we use a convergent expansion of the Whittaker function  $M_{\lambda,\mu}(z)$ in a
series of Bessel functions given by Buchholz \cite{Buch}. It reads
\begin{equation}
M_{\lambda,\mu}(z)=\Gamma(2\mu+1)2^{2\mu}z^{\mu+\frac{1}{2}}\sum\limits_{n=0}^\infty p_n^{(2\mu)}(z)\frac{J_{2\mu+n}(2\sqrt{\lambda z})}{(2\sqrt{\lambda z})^{2\mu+n}}\,,
\end{equation}
where $p_n^{(2\mu)}(z)$ are the Buchholz polynomials. Assume that $f(l)$ is a function such that $f(l)\simeq 1/l$ for $l>\!\!>1$ and let now $L>\!\!>1$ be an integer. Then we have for $z\to 0$
\begin{eqnarray}
&&\!\!\!\!\!\! \frac{1}{z}\sum\limits_{l=1}^\infty f(l)|M_{2l,\frac{1}{2}}(z)|^2\simeq\frac{1}{z}\sum\limits_{l=1}^{L=1}f(l)|M_{2l,\frac{1}{2}}(z)|^2+\frac{1}{z}\sum\limits_{l=L}^\infty\frac{1}{l}|M_{2l,\frac{1}{2}}(z)|^2\\
&&= C(L)z+4z^2\sum\limits_{l=L}^\infty\frac{1}{zl} \sum\limits_{m,n} p_n^{(1)}(z)p_m^{(1)}(z)
\frac{J_{1+n}(2\sqrt{2l z})J_{1+m}(2\sqrt{2l z})}{(2\sqrt{2l z})^{2+n+m}}\,,\nonumber
\end{eqnarray}
where $C(L)$ is a $z$-independent constant that depends logarithmically on $L$.
To continue we note that $p_n^{(1)}(z)$ is bounded by $z^n$ with $p_0^{(1)}(z)=1$. Furthermore for $z\to 0$ we can replace the sum over $l$ by an integral so that
\begin{eqnarray}
\frac{1}{z}\sum\limits_{l=1}^\infty f(l)|M_{2l,\frac{1}{2}}(z)|^2&\simeq& Cz\\&&+4z \sum\limits_{m,n} z^{n+m}\int\limits_{8zL}^\infty\frac{dx}{x}\frac{J_{2\mu+n}(\sqrt{x})J_{2\mu+m}(\sqrt{x})}{(\sqrt{x})^{2+n+m}}\,.\nonumber
\end{eqnarray}
This integral is well defined and finite apart from a logarithmic divergence for $x\to 0$.
The $L$-dependence between the first and second line cancels and we are left with
\begin{eqnarray}
\frac{1}{z}\sum\limits_{l=1}^\infty f(l)|M_{2l,\frac{1}{2}}(z)|^2&\simeq& 4z (\log(z)+const + ..)\,.
\end{eqnarray}

\section{Appendix 3}
To find the absorption cross section for a dilation with arbitrary angular momentum for vanishing D0-charge, $f=0$,  we consider
the dilation perturbation of the form
\begin{equation}
\delta\phi(t,\theta,\varphi)= Y_{LM}(\theta,\varphi)e^{ i \omega(t\mp\frac{R^2}{r})}\ .
\end{equation}
taking (\ref{hpw}), since now we have set of three spherical harmonic functions, the orthogonality condition we have used to write down (\ref{cao}) does not apply here, hence we have to keep summation over $l_1, m_1, l_2, m_2$ and for example $a^\dagger_{lm} a^\dagger_{l-m}$ replaces by $a^\dagger_{l_1 m_1} a^\dagger_{l_2 m_2}$ and so on. \\
The integral over sphare can be expressed by $3j~symbol$ as
\begin{eqnarray}
\int_{S^2} y^*_{m_1 l_1} y^*_{m_2 l_2} y_{LM} &=& (-1)^{l_1+l_2-L+M}\sqrt{\frac{(2l_1+1)(2l_2+1)(2L+1)}{4\pi}} \nonumber \\
&&\times \begin{pmatrix}
l_1&l_2&L\cr
0&0&0\cr
\end{pmatrix}\nonumber
\begin{pmatrix}
l_1&l_2&L\cr
m_1&m_2&-M\cr
\end{pmatrix}\nonumber
\end{eqnarray}
which, in turn can be calculated as a finite
sum by using the $Racah formula$ \cite{Messiah}
\begin{eqnarray}
&&\!\!\!\!\! \begin{pmatrix}
l_1&l_2&L\cr
m_1&m_2&-M\cr
\end{pmatrix}=(-1)^{l_1-l_2+M}\sqrt{\Delta(l_1l_2L)} \\
&&\!\!\!\!\! \times\sqrt{(l_1-m_1)!(l_1+m_1)!(l_2-m_2)!(l_2+m_2)!(L-M)!(L+M)!}\times\sum_t\frac{1}{f(t)}\nonumber
\end{eqnarray}
where $\Delta(l_1l_2L)$ is a triangle coefficient
\begin{equation}
\Delta(l_1l_2L):=\frac{(l_1+l_2-L)!(l_2+L-l_1)!(L+l_1-l_2)!}{(1+l_1+l_2+L)!}
\end{equation}
and
\begin{equation}
f(t):= t!(L-l_2+m_1+t)!(L-l_1-m_2+t)!(l_1+l_2-L-t)!(l_1-m_1-t)!(l_2+m_2-t)!\\
\end{equation}
\\
note that sum goes over all integer values of $t$ for which the arguments of factorials are non-negative.
\section{Appendix 4}
In this appendix we give a detailed calculation of the integral in (\ref{d00}). If we define a new variable $x$ through
\begin{equation}
t-\frac{1}{r}=\frac{\cosh\chi_0 \sin\tau-1}{\cosh\chi_0 \cos\tau+\sinh\chi_0}:=x\,,
\end{equation}
we have
\begin{eqnarray}
\int_{-\frac{\pi}{2}}^{\frac{3\pi}{2}} d\tau e^{i\omega(t-\frac{1}{r})} e^{-i N \tau}
&=&\int d x e^{i\omega x} e^{-iN \tau(x)}\frac{d\tau}{dx}\\
&=&2A\int dx e^{i\omega x}\frac{1}{1+x^2} \left(\frac{1-ix}{1+ix}\right)^N \nonumber \\
 &=& 2A\frac{2\pi i}{N!}(-i)^{N+1}\left[ e^{i\omega x} (1-ix)^{N-1}\right]^{(N)}_{x=i}\,, \nonumber
\end{eqnarray}
where $A=\left(\tanh\chi_0-\frac{i}{\cosh\chi_0}\right)^N$.
The result of integral is thus
\begin{eqnarray}
\int_{-\frac{\pi}{2}}^{\frac{3\pi}{2}} d\tau e^{i\omega(t-\frac{1}{r})} e^{-i N \tau}=\\
(2\pi)\left(i\frac{1}{\cosh\chi_0}-\tanh\chi_0\right)^N  \!\!\!\!\ && \!\!\!\!\!\!\!\!\!
\sum_{k=0}^{N-1} {N-1\choose k} \frac{1}{(N-k)!} (-2\omega)^{N-k}e^{-\omega}\,. \nonumber
\end{eqnarray}
This sum is of the form
\begin{equation}
\sum_{k=0}^{N}{N\choose k}\frac{1}{(N+1-k)!}(-2\omega)^{N+1-k}\,,
\end{equation}
with $N$ is $2l-1$, $2l$ or $2l+1$. Now we change the form of sum so that we can write it in terms of Hypergeometric function, ie.
\begin{eqnarray}
\cdots &=&(-2\omega) \sum_{k=0}^{N}\frac{N!}{(N-k)!(N+1-k)!k!}(-2\omega)^{N-k} \\
&=&(-2\omega) \sum_{p=0}^{N}\frac{N!}{(p)!(p+1)!(N-p)!}(-2\omega)^{p} ~~~~~~~~p:=N-k\nonumber \\
&=&(-2\omega) \sum_{p=0}^{N}\frac{(-N)_p}{(2)_p}\frac{(2\omega)^{p}}{p!} ~~~~(a=-N {~\text{and}~} b=p)\nonumber\\
&=&(-2\omega) M(-N,2,2\omega)=-e^{\omega}M_{1+N,\frac{1}{2}}(2\omega)\,,\nonumber
\end{eqnarray}
where $M(a,b,z)$
is Confluent Hypergeometric function and $M_{\lambda,\mu}(z)$ is Whittaker function.
\newpage

\end{document}